\documentclass[11pt]{article}

 \advance \textwidth by -2in
 \advance \textheight by -3in
\def\##1{\underline{#1}}
\def\=#1{\underline{\underline{#1}}}

\def\+#1{\underline{\bf #1}}
\def\*#1{\underline{\underline{\bf #1}}}

\def\eps{\epsilon}

\def\.{\mbox{ \tiny{$^\bullet$} }}

\def\curl{\nabla\times}
\def\div{\nabla \mbox{ \tiny{$^\bullet$} }}

\def\le{\left(}
\def\ri{\right)}

\def\lec{\left\{}
\def\ric{\right\}}

\def\c#1{\cite{#1}}

\def\r#1{(\ref{#1})}

\def\Dr{\#D(\#r,\omega)}
\def\Br{\#B(\#r,\omega)}
\def\Hr{\#H(\#r,\omega)}
\def\Er{\#E(\#r,\omega)}
\def\ro{(\#r,\omega)}

\def\rhoe{\rho_e}
\def\rhom{\rho_m}
\def\Je{{\#J}_e}
\def\Jm{{\#J}_m}

\begin{document}
\vskip 0.4cm

\noindent
{\bf A CONJUGATION SYMMETRY IN LINEAR ELECTROMAGNETISM, \\
IN
EXTENSION OF MATERIALS WITH
NEGATIVE REAL PERMITTIVITY AND PERMEABILITY SCALARS}
\vskip 0.2cm

\noindent  {\bf Akhlesh Lakhtakia}
\vskip 0.2cm

\noindent {\sf CATMAS~---~Computational \& Theoretical Materials Sciences Group \\
\noindent Department of Engineering Science \& Mechanics\\
\noindent 212 Earth \& Engineering Sciences Building\\
\noindent Pennsylvania State University, University Park, PA 16802--6812
}
\vskip 0.4cm

\noindent {\bf ABSTRACT:}
If all space is occupied by a linear
bianisotropic material~---~whether homogeneous
or not~---~then the concurrent replacements of the permittivity and
the impermeability tensors by the negatives of their respective
complex conjugates and of the magnetoelectric tensors by their respective
complex conjugates (in the Boys--Post representation) imply the conjugation of
both $\#E$ and $\#H$, in the absence of externally
impressed sources. This conjugation symmetry in linear
electromagnetism has observable
consequences when the linear bianisotropic material
occupies a bounded region.


\vskip 0.4cm

\noindent {\bf 1. INTRODUCTION}

The modest aim of this communication is to present a conjugation
symmetry of frequency--domain electromagnetic  fields in linear, nonhomogeneous,
 bianisotropic materials. This symmetry emerged as a generalization
 of a result obtained initially for linear, homogeneous, isotropic,
 dielectric--magnetic materials with negative real permittivity
 and permeability scalars \c{Ves}--\c{LMW}. Nominally, such a material
 possesses a relative permittivity scalar
$\eps_r =\eps_r' + i \eps_r''$ and a relative permeability scalar 
$\mu_r=\mu_r'+i\mu_r''$, both dependent
on the angular frequency $\omega$, such that both
 $\eps_r'<0$ and $\mu_r'<0$
 in some spectral
regime; accordingly, the phase velocity vector and the time--averaged
 Poynting vector are oppositely directed in that spectral regime \cite{LMW}.\footnote{The
condition for the phase velocity and the time--averaged Poynting vectors to be
oppositely directed is $\le\vert\eps_r\vert-\eps_r'\ri\le\vert\mu_r\vert-
\mu_r'\ri > \eps_r''\mu_r''$, which
 permits~---~more generally~---~$\eps_r'$
and/or  $\mu_r'$ to be negative \c{MLW}. An $\exp(-i\omega t)$
time--dependence having been assumed here, $\eps_r'' >0$
and $\mu_r'' >0$ at all $\omega>0$ for all passive
materials.}
Originally conceived  more than 35 years ago \c{Ves}, these materials
were artificially realized quite recently \c{SSS}.

During an investigation of changes in frequency--domain electromagnetic fields 
when the transformation $\left\{ \eps_r'\to-\eps_r',\,\right.$ $\left. \mu_r'\to-\mu_r'\right\}$
is effected on the isotropic dielectric--magnetic material occupying
a source--free region, a more general conjugation symmetry in linear
electromagnetism began to take shape. The following sections of this
communication report the development of that symmetry.

\noindent{\bf 2. CONJUGATE INVARIANCE OF MAXWELL POSTULATES}

The frequency--domain Maxwell postulates 
may be written as
\begin{equation}
\label{MP}
\left.
\begin{array}{l}
\div \Dr = \rhoe\ro\\

\div \Br = \rhom\ro\\
\curl\Er= i\omega \Br-\Jm\ro\\

\curl\Hr=- i\omega \Dr+\Je\ro
\end{array}\right\}\,,
\end{equation}
in the presence of externally impressed sources of the electric and magnetic types.
These four postulates are collectively  invariant with respect to the transformation
\begin{equation}
\label{ciMP}
\begin{array}{l}
\left\{ \#E\to \#E^\ast,\, \#H\to\#H^\ast,\,\#D\to -\#D^\ast,\, \#B\to-\#B^\ast,\,\right.
\\
\left.\quad
\Je\to\Je^\ast,\,\Jm\to\Jm^\ast,\,\rhoe\to-\rhoe^\ast,\,\rhom\to-\rhom^\ast\right\}\,,
\end{array}
\end{equation}
where the asterisk denotes the complex conjugate.
This statement of conjugate invariance may be verified by direct substitution
of \r{ciMP} in \r{MP}. 

The  conjugate
invariance of the Maxwell postulates not only underlies
a similarly invariant Beltrami form of electromagnetism \c{BMPco}, but also
permits the existence of 
a conjugation symmetry in linear electromagnetism.

\newpage

\noindent{\bf 3. CONJUGATE INVARIANCE AND LINEAR BIANISOTROPY}

There are two widely used sets of frequency--domain
constitutive relations for linear bianisotropic
materials \c{Wei}. Both are considered as follows:
\begin{itemize}
\item {\em Boys--Post constitutive relations.\/}
The Boys--Post constitutive relations of a linear, nonhomogeneous, bianisotropic material
can be stated as 
\begin{equation}
\label{conrelBP}
\left.
\begin{array}{l}
\Dr = \=\eps\ro\.\Er + \=\alpha\ro\.\Br\\
\Hr = \=\beta\ro\.\Er + \=\chi\ro\.\Br
\end{array}\right\}\,,
\end{equation}
subject to the constraint ${\rm Trace}\lec \=\alpha-\=\beta\ric\equiv0$.
Whereas $\=\eps$ is the permittivity tensor
and $\=\chi$ is the impermeability tensor, both $\=\alpha$ and
$\=\beta$ are magnetoelectric tensors.\\

\bigskip

The transformation
\begin{equation}
\label{tfmBP}
\lec \=\eps\to-\=\eps^\ast,\, \=\chi\to-\=\chi^\ast,\,\=\alpha\to\=\alpha^\ast,\,
\beta\to\=\beta^\ast\ric
\end{equation}
of constitutive tensors
then entails the transformation
\begin{equation}
\label{fBP}
\lec \#E\to \#E^\ast,\, \#H\to\#H^\ast,\,\#D\to -\#D^\ast,\, \#B\to-\#B^\ast\ric
\end{equation}
of the electromagnetic fields~---~in conformity with  \r{ciMP}
expressing the conjugate invariance of the Maxwell postulates.

\item {\em Tellegen constitutive relations.\/}
The Tellegen constitutive relations of a linear, nonhomogeneous, bianisotropic material
can be stated as
\begin{equation}
\label{conrelT}
\left.
\begin{array}{l}
\Dr = \={\hat\eps}\ro\.\Er + \=\xi\ro\.\Hr\\
\Br = \=\zeta\ro\.\Er + \=\mu\ro\.\Hr
\end{array}\right\}\,,
\end{equation}
subject to the constraint ${\rm Trace}\lec\=\mu^{-1}\.(\=\xi+\=\zeta)\ric\equiv0$.
Here, $\={\hat\eps}$ is the permittivity tensor,
$\=\mu$ is the permeability tensor, and both $\=\xi$ and
$\=\zeta$ are magnetoelectric tensors.\\

\bigskip

The transformation
\begin{equation}
\label{tfmT}
\lec \={\hat\eps}\to-\={\hat\eps}^\ast,\, \=\mu\to-\=\mu^\ast,\,\=\xi\to-\=\xi^\ast,\,
\zeta\to-\=\zeta^\ast\ric
\end{equation}
of constitutive tensors
then also entails the field transformation \r{fBP}
in conformity with  the conjugate invariance of the Maxwell postulates.

\end{itemize}
Because the frequency--domain constitutive relations
of the Boys--Post and the Tellegen types
are inter--translatable, the constitutive--tensor  transformations
\r{tfmBP} and \r{tfmT} are actually equivalent.

\noindent{\bf 4. CONJUGATION SYMMETRY IN LINEAR ELECTROMAGNETISM}

The deductions in Section 3 permit the
enunciation of the following
{\em conjugation symmetry}:
If all space were to be occupied by a linear
bianisotropic material~---~whether homogeneous
or not~---~then a change of the constitutive tensors
as per \r{tfmBP} would imply the conjugation of
both $\Er$ and $\Hr$, in the absence of externally
impressed sources. If such sources are present, then the conjuagtion
symmetry is expressed jointly by \r{ciMP} and
\r{tfmBP}.

The effect of the transformation
\r{tfmBP} would be observable even if the linear bianisotropic
material were to be confined to a bounded region.
For instance, imagine a  slab of infinite transverse extent and
uniform thickness, separating two vacuous half--spaces;
the slab is made of a linear, isotropic, dielectric--magnetic
material; and a linearly polarized, propagating, plane wave is incident on the slab. If the real
parts of the permittivity and the permeability scalars
of the slab were to change sign simultaneously,
then the reflection and the transmission coefficients
would have to be replaced by their respective complex
conjugates \c{Lakh1}. The same conclusion holds
if the slab were to be piecewise uniform in the thickness direction. The 
complex conjugation
of reflection and transmission coefficients would essentially
hold even if the slab were to be bianisotropic and
plane--stratified, but a dependence on the polar
angle of the incidence wavevector would also appear~---~as
demonstrated elsewhere
for chiral ferrosmectic slabs \c{Lakh2}.
However, the conjugation of the reflection and the transmission
coefficients would not hold on reversal of the signs
of the real parts of the permittivity and permeability scalars
of an isotropic, plane--stratified, dielectric--magnetic slab, if the incident
plane wave were to evanescent.

The conclusions stated in the foregoing paragraph were obtained
both by direct calculation and by application of the conjugation symmetry
enunciated at the beginning of this section. The latter procedure is,
of course, very simple; and it demonstrates the importance of
the proffered symmetry in quickly determining the
observable consequences of employing newly emerging
materials (such as isotropic dielectric--magnetic
materials with negative real permittivity and permeabilty scalars, and
their anisotropic counterparts) for various applications.

\end{document}